# Screening of antibacterial activity of lactic acid bacteria isolated from fermented vegetables against food borne pathogens

Mahnaz Kazemipoor[1] (Corresponding author)

Department of Science & Technology Studies,

Faculty of Science, University of Malaya,

Kuala Lumpur, Malaysia

Tel: +60-10-547-2074   E-mail: Mahnaz@siswa.um.edu.my

Che Wan Jasimah Wan Mohamed Radzi[2]

Department of Science & Technology Studies,

Faculty of Science, University of Malaya,

Kuala Lumpur, Malaysia

Tel: +60-37-967-5182   E-mail: jasimah@um.edu.my

Khyrunnisa Begum[3]

Department of Studies in Food Science and Nutrition,

University of Mysore, Mysore, India

Tel: +91-821-2419361  E-mail: akhilaftab@yahoo.com

Iman Yaze[4]

Department of Physiology,

Faculty of Medicine, University of Malaya,

Kuala Lumpur, Malaysia

Tel: +60-14-914-5737   E-mail: iman.yaze@yahoo.com

**Abstract**

This study aims to screen the antibacterial activity of lactic acid bacteria (LAB) isolated from home-made fermented vegetables against common food borne pathogens. The antagonistic properties of these isolates against Escherichia coli, Staphylococcus aureus, Yersinia enterocolitica and Bacillus cereus were examined using agar well diffusion method. Four LAB namely MF6, MF10, MF13, and MF15 identified as Lactobacillus animalis, Lactobacillus rhamnosus, Lactobacillus fermentum and Lactobacillus reuteri, respectively were effective against all selected pathogenic strains. Amongst the four isolates, MF6 exhibited the highest antibacterial activity, against all the indicator pathogens tested except Y. enterocolitic. Its activity was maximum against E.coli with a Zone of Inhibition (ZOI) ranging from 18.7 to 21.3 mm and least for Y. enterocolitica (10±1.1 mm). Isolate MF13 also showed antimicrobial property against all tested pathogens showing highest activity against Y. enterocolitica (14 ± 1.7 mm) and least against E.coli (8 ± 1.4 mm), which was in direct contrast to isolate MF6. Isolate MF15 showed greater activity against E.coli (12 ± 0.8 mm) and least against S. aureus (8 ± 1.7 mm). Least antimicrobial property was observed in isolate MF10, with a ZOI in the range of 2.5-7 mm. The degree of antimicrobial property among the isolates was in the order of MF6>MF13>MF15>MF10. Overall, the isolated LAB showed the remarkable inhibitory effect against both Gram positive and Gram negative pathogenic strains. However, the spectrum of inhibition was different for the isolates tested. These results suggest that this potent isolates could be used as a natural biopreservatives in different food products.

## 1. Introduction

Nowadays, food is no longer considered by consumers only in terms of taste and immediate nutritional needs, but also in terms of their ability to provide specific health benefits beyond their basic nutritional value. Currently, the largest segment of the functional food market is dominated by healthy food products targeted towards improving the balance and activity of the intestinal microflora [1]. Consumption of food containing live bacteria is the oldest and still most widely used way to increase the number of advantageous bacteria called "probiotics" in the intestinal tract [2]. Noteworthy, there are a large number of probiotic foods which date back to ancient times which are mostly originated from fermented foods as well as cultured milk products [2-10] The quest to find food ingredients with valuable bioactive properties has encouraged interest in lactic acid bacteria (LAB) with probiotic attributes such as antimicrobial activity against pathogenic microorganisms [11], antiviral activity [12], anti-yeast property [13], antimutagenic [14], antiplatelet aggregation [15], and antioxidant attributes [16] etc.

In general, it is believed that probiotics help keep up the balance between harmful and beneficial bacteria in the gut thus maintaining a healthy digestive system [10]. The health claims of probiotics range from regulation of bowel activity and well-being to more specific actions such as, antagonistic effect on the gastroenteric pathogens like Clostridium difficile, Campylobacter jejuni, Helicobacter pylori and Rotavirus etc [17]. Some are known to neutralize food mutagens produced in colon, shifting the immune response towards a Th2 response, alleviating allergic reactions, and lowering serum cholesterol [18].

The mechanism of action of probiotics with anti-microbial properties is maybe due to the production of bacteriocins such as nicin [19] or lowering the pH by producing acidic compounds like lactic acid, [20]. Probiotic strains compete with other infectious bacteria for nutrients and cell-surface and help toward them off by inhibiting their colonization [21]. A few strains are also known to produce active enzymes which inhibit other pathogenic bacteria [22]. The health benefits of probiotics have always been investigated with regard to their capability to sustain their availability, viability [23], digestibility, and rendering of their health benefits to the host without altering the safety [24] and the organoleptic properties of the food in which they have been incorporated [25]. Today, viable probiotic strains with beneficial functional properties are available in the market as components of foods and beverages, in fermented dairy products like yogurt [26] or as probiotic fortified foods as well as food preservatives [27].

## 2. Materials and Methods

### 2.1 Samples for isolation of LAB

The samples for the isolation of LAB included the fermented mixed vegetables which was prepared by mixing thoroughly, cleaned and diced vegetables (carrot, green beans, snake gourd, eggplant, green chili, bitter gourd, turnip, beet root, ladies finger, radish, cabbage) in 2% salt solution, and was allowed to ferment at ambient temperature (32°C±2), for a period of two months.

### 2.2 Media and reagents

Isolation as well as culturing of LAB was done using the De Man Rogosa Sharpe (MRS) media and the potato dextrose agar (PDA) media for yeasts. For the antimicrobial assay, the pathogenic cultures namely E. coli, S. aureus, Y. enterocolitica and B. cereus were grown in brain heart infusion (BHI) agar medium. The nutrient agar (NA) media was also used for the antimicrobial assay. The reagents, indicator, carbon substrates for biochemical tests etc. used in the study were of analytical grade and procured from Hi Media Chemicals Ltd., Mumbai, India.

### 2.3 Isolation of LAB

Isolation of LAB from the selected sample was carried out using the microbial pour plate method. Appropriate dilutions of the samples prepared in saline (0.85% NaCl) were pour plated into MRS medium, for isolation of LAB and PDA for isolation of yeast. The plates were then incubated at 37oC±2oC, for a period of 24-48 hrs. Colonies which were different from each other in their morphology and phenotypic appearance were picked up and inoculated in agar slants. Colonies which were sub-surface were inoculated as stab.

The presumptive LAB as well as yeast isolates were purified using their respective isolation media by re-streaking on plates until only a single type of colony was present. The different pure cultures so obtained were characterized for their colony morphology and subjected to Gram staining. Only Gram positive, non-motile, rod shaped bacteria, showing phenotypic characters similar to Lactobacillus species on MRS agar media (MF6, MF10, MF13, and MF15) were selected for further experiments. The cultures were stored and maintained at 4°C on MRS agar slants/stabs for further studies.

### 2.4 Determination of anti-microbial activity of MF6, MF10, MF13 and MF15

### 2.4.1 Preparation of sample filterate

The selected LAB isolates (MF6, MF10, MF13 and MF15) were inoculated from slants to fresh 250 ml MRS broth and incubated at 37°C for 48 hrs. The culture broth of each isolate was centrifuged separately at $10,000 \times g$ (Sorvall super-speed RC2-B) for 30 minutes. The supernatant was collected after centrifugation and passed through 0.2 µm sterile syringe filter (Fisher Scientific Co., Fair Lawn, NJ). To confirm bacteriocin production, the cell free neutral supernatant broths was collected for the antibacterial study against selected food borne pathogens.

### 2.4.2 Media for growth of pathogens

The pure cultures of food borne pathogens namely E. coli, S. aureus, Y. enterocolitica and B. cereus were inoculated from slants to brain heart infusion broth (BHIB). After 24 hr incubation at 37°C, the culture broth was centrifuged and the pellet obtained was suspended in 9 ml saline. This suspension was used for inoculation of the pathogenic strain to nutrient agar plates for the antimicrobial activity determination of the sample filterate.

### 2.5 Antimicrobial activity test by agar well diffusion method

The agar well diffusion method was used to determine the antimicrobial property of the LAB isolates. A 24 hr culture of the pathogens (E. coli, S. aureus, Y. enterocolitica and B. cereus), grown in BHIB at 37°C was suspended in saline. A lawn of the indicator strain was made by spreading the cell suspension over the surface of nutrient agar plates with a sterile cotton swab. The plates were allowed to dry and a sterile cork borer of diameter (5 mm) was used to cut uniform wells in the agar. Each well was filled with 60 µl culture free filterate obtained from the LAB isolates. After incubation at 37°C for 48 hrs, the plates were observed for a zone of inhibition (ZOI) around the well. Results were considered positive if the diameter (mm) of the ZOI was greater than 1mm. The experiment was carried out in triplicates and activity was reported as diameter of ZOI ± SD.

### 2.6 Identification of LAB isolates

The identification of potent isolates upto species level was done based on the characteristics of Lactobacillus as described in Bergey's Manual of Systematic Bacteriology [28] and a descriptive table given by Nair and Surendran [29]. The cultures were subjected to a battery of biochemical tests which included fermentation of different carbon sources, acid and gas production from glucose, catalase test, growth at different temperatures (15°C, 45 °C and both) and hydrolysis of arginine.

### 3. Results and Discussion

### 3.1 Isolation of lactic acid bacteria (LAB) from fermented vegetables

Lactic acid bacteria (LAB) and yeast cultures were isolated from fermented vegetables. All the isolates obtained were morphologically characterized. Table 1 gives the colony characteristics of the isolates obtained along with their Gram reaction and microscopic examination. Only Gram positive, non-motile, rod shaped bacteria, showing phenotypic characters similar to Lactobacillus species on MRS agar media (MF6, MF10, MF13, and MF15) were selected for further experiments. The basis for selection was that these isolates could probably belong to the Genus Lactobacillus, which has been shown to have probiotic attributes [30].

### 3.2 Determination of the antimicrobial activity of selected LAB by agar well diffusion method

The agar well diffusion method was used to assess the antimicrobial activity of the selected LAB namely MF6, MF10, MF13, and MF15 isolated from varied sources. Their antimicrobial properties were tested against four major food-borne pathogenic bacteria namely E. coli, S. aureus, Y. enterocolitica and B. cereus. Results show that the spectrum of inhibition was different for the isolates tested. Table 2 gives the results for the antimicrobial activity of the isolates in terms of diameter of the zone of inhibition (ZOI). A diameter >1mm around the well was considered as a positive result. It was assumed that greater the diameter of the ZOI, greater was the antimicrobial activity of the isolate. Plate 1 shows a petri plate with a lawn of an indicator pathogenic strain and a ZOI around the well containing the culture-free filterate.

Results indicate that isolate MF6 had the highest antimicrobial property, against all the indicator pathogens tested except Y. enterocolitica, amongst the four isolates. Its activity was highest against E.coli with a ZOI of 20 ± 1.3 mm and least for Y. enterocolitica (10±1.1 mm). Isolate MF13 also showed antimicrobial property against all tested pathogens with its activity being highest against Y. enterocolitica (14 ± 1.7 mm) and least against E.coli (8 ± 1.4 mm), which was in direct contrast to isolate MF6. Isolate MF15 showed greater activity against E.coli (12 ± 0.8 mm) and least against S. aureus (8 ± 1.7 mm). Least antimicrobial activity against all the tested indicator pathogenic bacteria was observed in isolate MF10, with a ZOI in the range of 2.5-7 mm. The degree of antimicrobial property among the isolates was in the order of MF6>MF13>MF15>MF10. It is however difficult to comment on the reason for this variability in the antimicrobial property amongst the isolates since each one was different from the other. Interestingly, isolate MF6 and MF13 were equally antagonistic against both gram positive (Staphylococcus aureus, Bacillus cereus) as well as gram negative (Escherichia coli and Yersinia enterocolitica) pathogenic bacteria.

### 3.3 Identification of LAB isolates

Table 3 gives the results of the various biochemical tests performed on the selected isolates. The most potent probiotic isolate MF6 which showed the highest antimicrobial property against all the food borne pathogens tested was identified as Lactobacillus animalis. The isolate MF13 was identified as L.fermentum. During the biochemical tests, isolate F10 and F15 were found to be weakly catalase positive, however, in spite of this, they were subjected for biochemical tests since many native Lactobacillus isolates have been previously reported to possess pseudocatalase activity [31-33]. Isolate F10 and F15 may be L. rhamnosus, and L. reuteri respectively. However, 16S rDNA analysis will be required to identify the isolates correctly.

## 4. Conclusion

The inhibitory action of LAB bacteria can be due to the accumulation of main primary metabolites such as lactic and acetic acids, ethanol and carbon dioxide. Additionally, LAB are also capable of producing antimicrobial compounds such as formic and benzoic acids, hydrogen peroxide, diacetyl, acetoin and bacteriocins such as nicin [19]. The production levels and the proportions among those compounds depend on the strain, medium compounds and physical parameters [34]. The inhibitory activities of LAB against Gram positive pathogens have been mostly shown to be due to the bactericidal effect of protease sensitive bacteriocins [35]. However, the antagonistic effects of LAB towards Gram negative pathogens could be related to the production of organic acids and hydrogen peroxide [36].

In conclusion, the results obtained from this study demonstrated the remarkable antimicrobial attributes of the isolated lactobacillus species from fermented mixed vegetables. According to previous studies, a large number of lactic acid bacteria strains with different bioactive potentials especially in the form of antimicrobial properties have been identified from a variety of plant sources mostly in the form of fermented and pickled vegetables [37-42]. These scientific evidences have been a motivating factor to choose a plant based fermented product prepared from different vegetables which could further confirm the results of this study. On the other hand, such positive outcomes would be a leading point towards application of simple worthy traditional methods such as fermentation in producing natural healthy food products and encouraging consumers to include such valuable food items into their eating habits. Hope these friendly food groups would be added to daily diet of each individual to improve their body immunity hence, decreasing unnecessary intake of chemical antibiotics. However, further in vitro and in vivo studies are required according to selection criteria including adhesion to mucosal cells of the gastrointestinal tract, bile salt and acid tolerance, bile salt hydrolase activity, viability, resistance to antibiotics, safety and organoleptic properties to be applicable in different food products such as starter culture in fermented dairy products.

**List of abbreviations:**

LAB     Lactic Acid Bacteria

ZOI     Zone of Inhibition

MRS     De Man Rogosa Sharpe media

PDA     Potato Dextrose Agar media

BHI     Brain Heart Infusion Agar media

NA      Nutrient Agar media

**Acknowledgment:**

The authors wish to thank Institute of Research Management & Monitoring IPPP and Department of Sceicne & Technology Studies, Faculty of Science, University of Malay for assistance in page charge fund for publication of this paper.

Table 1- Colony morphology of cultures isolated from different sources

| No. | Shape | Size | Margin | Elevation | Pigmentation | Gram stain |
|---|---|---|---|---|---|---|
| MF6 | Circular | Small | Entire | Convex | White | G+ long rods in chains |
| MF10 | Circular | Small | Entire | Convex | Cream | G+ rods in pairs |
| MF13 | Circular | Small | Entire | Convex | White | G+ rods in chains |
| MF15 | Circular | Small | Entire | Raised | Cream | G+ rods in chains |

Table 2- Antimicrobial activity of the isolates in terms of ZOI using the agar well diffusion method

| Pathogen | ZOI (mm±SD) |
|---|---|

|  | MF6 | MF10 | MF13 | MF15 |
|---|---|---|---|---|
| *B. cereus* | 13±1.8 | 3±1.7 | 12±1.5 | 10±1.4 |
| *Y. enterocolitica* | 10±1.1 | 7±1.4 | 14±1.7 | 11±1.3 |
| *S. aureus* | 13±1.7 | 3±2.6 | 11±2.4 | 8±1.7 |
| *E. coli* | 20±1.3 | 2.5±1.6 | 8±1.4 | 12±0.8 |

Table 3- Results of the biochemical tests carried out on the selected isolates (MF6, MF10, MF15 and MF13)

| Parameter | MF6 | MF10 | MF13 | MF15 |
|---|---|---|---|---|
| Growth at different temperatures |  |  |  |  |
| 15 °C only | - | - | + | + |
| 45 °C only | - | - | + | + |
| 15 & 45 °C | - | - | + | + |
| Acid & gas from glucose | +/- | +/- | +/+ | +/+ |
| NH$_3$ from Arginine | + | - | + | + |
| Sugar fermentation |  |  |  |  |
| Arabinose | + | - | + | + |
| Cellobiose | + | + | + | - |
| Mannitol | + | + | + | - |
| Mannose | + | + | + | - |
| Melebiose | + | - | + | + |
| Raffinose | + | - | + | - |

| Ribose | - | - | + | + |
|---|---|---|---|---|
| Salicin | + | + | + | - |
| Rhamnose | + | - | - | - |
| Xylose | - | + | + | + |
| Species identified | *L. animalis* | *L.rhamnosus* | *L.fermentum* | *L. reuteri* |

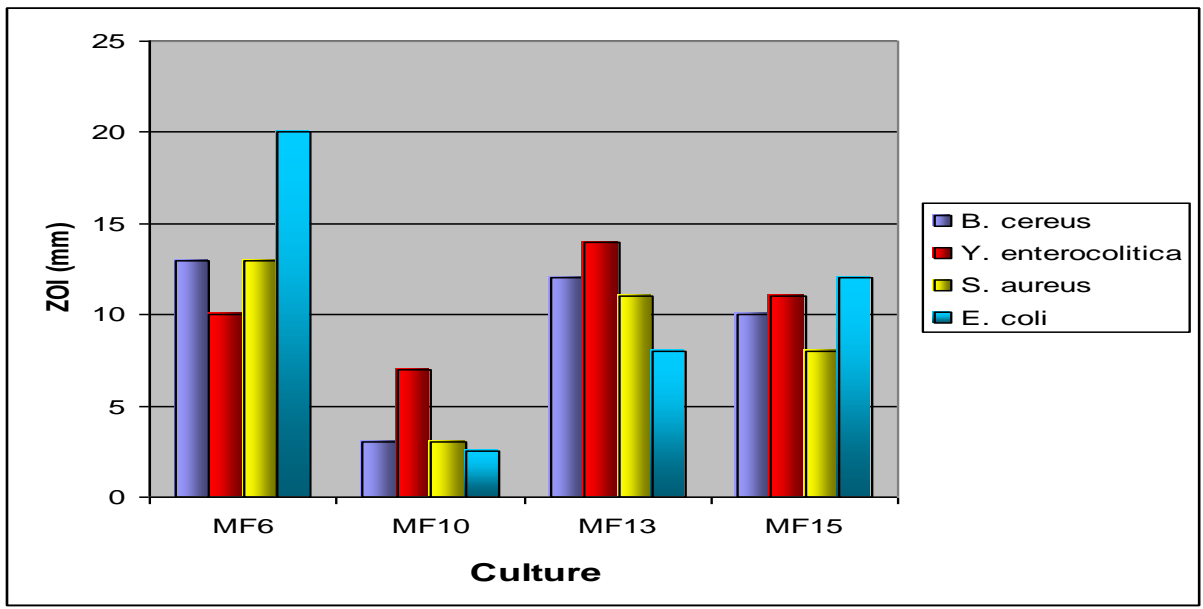

Fig. (1)- Antimicrobial activity of the isolates against selected food borne pathogens using the agar well diffusion method

**A**                                    **B**                                    **C**

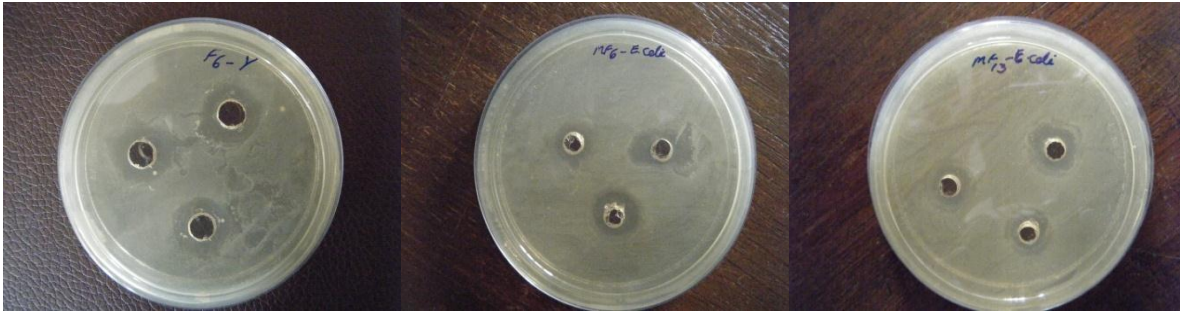

Plate 1: Petri Plate with a Lawn of an Indicator Pathogenic Strain and a Zone of Inhibition around the Well
Containing the Culture-Free Filterate of MF6 and MF13

A) MF6 against Y. enterocolitica        B) MF6 against E. coli                C) MF13 against E. coli